\documentclass[RNAAS]{aastex62}

\begin{document}

\title{A compendium of extinction curves for simple galactic geometries}

\correspondingauthor{Andrew Benson}
\email{abenson@carnegiescience.edu}

\author[0000-0001-5501-6008]{Andrew Benson}
\affiliation{Carnegie Institution for Science \\
813 Santa Barbara Street, Pasadena, CA 91101, USA}


\section{} 

Accounting for the effects of dust extinction on the observable properties of galaxies is a critical component in galaxy modeling and in constructing synthetic galaxy surveys. \cite{ferrara_atlas_1999}\footnote{See also \protect\cite{tuffs_modelling_2004} who provide fitting functions to the attenuation in similar models, and also consider clumping of the dust, and \protect\cite{fontanot_evaluating_2009} who provide a fit to the results of a radiative transfer code applied to galaxies from a semi-analytic galaxy formation model.} computed extinction curves for model galaxies with simple, yet realistic geometries, as a function of V-band central optical depth (measured through the center of the galaxy viewed face-on, $\tau_\mathrm{V,0}=\int_{-\infty}^{+\infty} \mathrm{d}z \rho(R,z) \kappa_\mathrm{V}$, where $\rho(R,z)$ is the dust density in cylindrical coordinates, and $\kappa_\mathrm{V}$ is the opacity of the dust in the V-band), inclination, and morphology. Such tabulations are useful for computing extincted magnitudes in galaxy formation models \citep{lacey_unified_2016,merson_predicting_2018}.

The demands of current and forthcoming survey simulation efforts require higher precisions \citep{mao_descqa:_2018}. The extinction curves of \cite{ferrara_atlas_1999} are tabulated at a relatively small number of points. Applying these extinction curves requires interpolation and extrapolation which we have found can lead to noticeable numerical artifacts in the resulting galaxy colors.

We have therefore conducted calculations similar to those of \cite{ferrara_atlas_1999}, using updated dust properties, spanning a wider range of inclination, optical depth, and morphology, and using a much larger number of tabulation points. We use the radiative transfer code {\sc Hyperion} \citep{robitaille_hyperion:_2011}, and dust properties corresponding to the model of \cite{draine_interstellar_2003} with intrinsic $R_\mathrm{V}=4.0$. The dust is distributed with density profile $\rho_\mathrm{d}(R,z) \propto \exp(-R/R_\mathrm{d}) \mathrm{sech}^2(z/z_\mathrm{d})$, where $R_\mathrm{d}$ and $z_\mathrm{d}$ are radial and vertical scale lengths respectively. Disk stars are distributed to trace the dust, while spheroid stars follow a \cite{hernquist_analytical_1990} profile with scale length $r_\mathrm{s}$. When characterized by $\tau_\mathrm{V,0}$ the attenuations depend only upon $z_\mathrm{d}/R_\mathrm{d}$ and $R_\mathrm{s}/R_\mathrm{d}$.

\begin{figure}[!ht]
  \begin{center}
    \includegraphics[scale=0.6,angle=0]{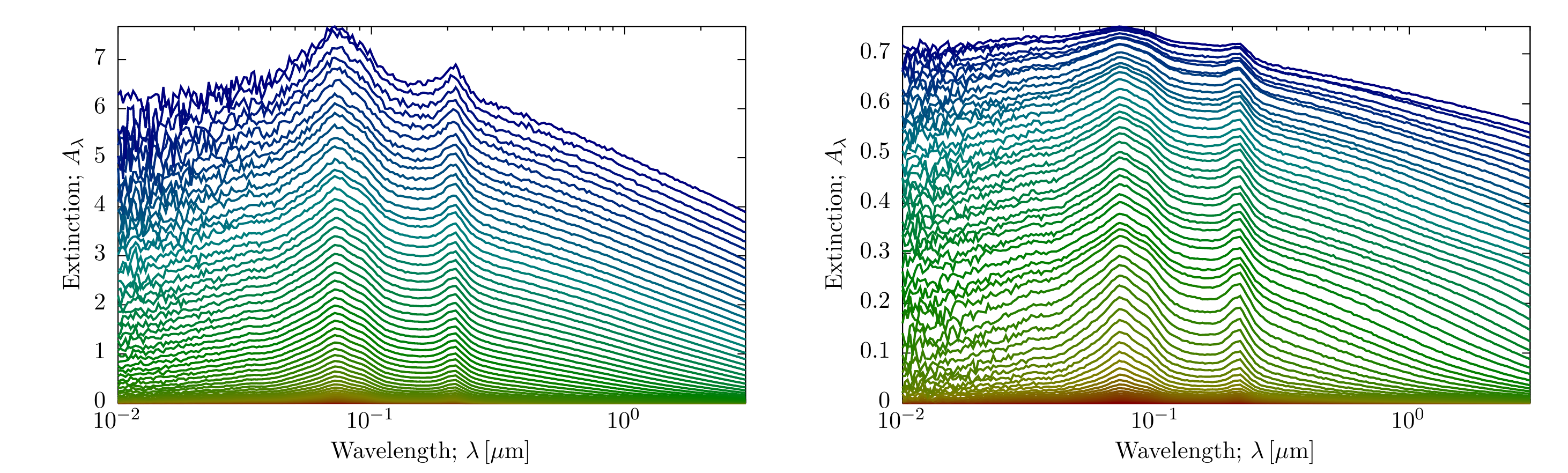}
    \caption{Attenuation curves, in magnitudes of extinction, for a model with $r_\mathrm{s}/r_\mathrm{d}=1$ at inclination angle $60^\circ$. Left and right panels show the attenuation for disk and spheroid components. Line color indicates $\tau_\mathrm{V,0}$ from $10^{-2}$ (red) to $10^4$ (blue).\label{fig:attenuation}}
  \end{center}
\end{figure}

Density profiles are tabulated on a $(100,100)$ grid in $(R,z)$, spaced in the logarithm of $R$ from $10^{-2}R_\mathrm{d}$ to $R_\mathrm{m}$, and linearly in $z$ between $\pm z_\mathrm{m}$. Here $R_\mathrm{m}=10\, \mathrm{max}(r_\mathrm{s},r_\mathrm{d})$ and $z_\mathrm{m}=10\, \mathrm{max}(r_\mathrm{s},z_\mathrm{d})$ are chosen to encompass the majority of the mass. We configure \textsc{Hyperion} to use monochromatic radiative transfer via raytracing, and to use $10^5$ photons for both ray tracing and imaging. We tabulate the attenuation $a_\lambda = L_{\mathrm{obs},\lambda}/L_{\mathrm{true},\lambda}$ at a large number of viewing angles, where $L_{\mathrm{obs},\lambda}$ is the observed luminosity at wavelength $\lambda$, and $L_{\mathrm{true},\lambda}$ is the true (unextincted) luminosity at that same wavelength. Due to scattering $a_\lambda > 1$ is possible. Example results are shown in Figure~\ref{fig:attenuation}.

The tabulated attenuation curves are made available as an HDF5 file, containing the following datasets:
\begin{description}
\item[\texttt{wavelength[N$_\lambda$]}] wavelength [microns];
\item[\texttt{opticalDepth[N$_\tau$]}] optical depth, $\tau_\mathrm{V,0}$;
\item[\texttt{inclination[N$_i$]}] inclination of the disk [degrees];
\item[\texttt{spheroidScaleRadial[N$_r$]}] radial scale length of the spheroid, relative to that of the disk [$r_\mathrm{s}/r_\mathrm{d}$];
\item[\texttt{attenuationDisk[N$_\lambda$,N$_i$,N$_\tau$]}] attenuation of the disk;
\item[\texttt{attenuationSpheroid[N$_\lambda$,N$_i$,N$_\tau$,N$_r$]}] attenuation of the spheroid;
\item[\texttt{attenuationUncertaintyDisk[N$_\lambda$,N$_i$,N$_\tau$]}] uncertainty in the attenuation of the disk;
\item[\texttt{attenuationUncertaintySpheroid[N$_\lambda$,N$_i$,N$_\tau$,N$_r$]}] uncertainty in the attenuation of the spheroid.
\end{description}
The following attributes are present in the file:
\begin{description}
  \item[\texttt{dustDescription}] a description of the dust model used;
  \item[\texttt{opacity}] the opacity of the dust at a wavelength of $0.55\mu$m [cm$^2$/g];
  \item[\texttt{diskStructureVertical}] a description of the vertical density profile of the disk;
  \item[\texttt{diskCutOff}] the radius (in units of the disk radial scale length) at which the computational grid was truncated;
  \item[\texttt{spheroidCutOff}] the radius (in units of the spheroid radial scale length) at which the computational grid was truncated;
  \item[\texttt{diskScaleVertical}] the vertical scale length of the stellar disk, relative to the radial scale length of the disk;
  \item[\texttt{dustScaleVertical}] the vertical scale length of the dust disk, relative to the radial scale length of the disk;
  \item[\texttt{timeStamp}] a timestamp indicating when the file was created.
\end{description}
This file is available from at \dataset[10.5281/zenodo.1442826]{https://doi.org/10.5281/zenodo.1442826}. We expect to add further cases (different dust and stellar scale heights, dusty spheroids, clumpy dust, etc.) in future. The \href{https://bitbucket.org/galacticusdev/analysis-python/src/master/}{\texttt{analysis-python}} package for \textsc{Galacticus} \citep{benson_galacticus:_2012} includes a module to apply these attenuation curves to \textsc{Galacticus} model outputs.

\acknowledgments

We thank Thomas Robitaille for developing the {\sc Hyperion} code.

\bibliography{compendium}

\end{document}